\documentclass[twocolumn]{AAS77860R3}

\usepackage{amsmath}
\usepackage{graphicx}
\usepackage{booktabs}
\usepackage{multirow}
\usepackage{placeins}
\usepackage{xcolor}

\newcommand{\fermi}{\textit{Fermi}-LAT}

\begin{document}

\title{A GeV PWN Candidate in the PSR J1135--6055/G293.8+0.6 Field}

\author{Liancheng Zhou}
\affiliation{Department of Astronomy, School of Physics and Astronomy, Key Laboratory of Astroparticle Physics of Yunnan Province, Yunnan University, Kunming 650091, China}
\affiliation{Southwest United Graduate School, Kunming 650092, China}
\email{lizhang@ynu.edu.cn}

\author{Yunlu Gong}
\affiliation{Department of Astronomy, School of Physics and Astronomy, Key Laboratory of Astroparticle Physics of Yunnan Province, Yunnan University, Kunming 650091, China}
\email{lizhang@ynu.edu.cn}

\author{Jun Fang}
\affiliation{Department of Astronomy, School of Physics and Astronomy, Key Laboratory of Astroparticle Physics of Yunnan Province, Yunnan University, Kunming 650091, China}
\email{junfang@ynu.edu.cn}

\author{Li Zhang}
\affiliation{Department of Astronomy, School of Physics and Astronomy, Key Laboratory of Astroparticle Physics of Yunnan Province, Yunnan University, Kunming 650091, China}
\email{lizhang@ynu.edu.cn}

\correspondingauthor{Li Zhang: lizhang@ynu.edu.cn}

\begin{abstract}
We present a \textit{Fermi} Large Area Telescope (LAT) study of PSR~J1135--6055 and the composite supernova remnant (SNR) G293.8+0.6.
 MeerKAT radio observations reveal a bright central plerion within the faint SNR shell, while Chandra observations detect a compact nebula around the pulsar. Together, these observations establish the presence of a PWN within G293.8+0.6. Our analysis of the 20~GeV--1~TeV LAT data yields a point-like excess with $\mathrm{TS}=26.30$ and photon index $\Gamma=1.63\pm0.33$. Its best-fit position is consistent with the pulsar within the 95\% statistical localization region, and neither the uniform-disk nor Gaussian model is favored, providing no evidence for significant extension. A one-zone leptonic model, together with an independent jet component, reproduces the multiwavelength observations. In this model, the GeV photons are produced mainly by inverse-Compton scattering from relativistic electrons and positrons. Given its positional agreement with the pulsar/PWN and broadband interpretation, we regard the LAT point source as a GeV PWN candidate.
\end{abstract}


\section{Introduction}
\label{sec:intro}
Pulsar wind nebulae (PWNe) and supernova remnants (SNRs) are important Galactic particle accelerators. In PWNe, relativistic electrons and positrons supplied by the pulsar wind produce synchrotron emission from radio to X-ray energies and can generate gamma rays through inverse-Compton scattering on ambient photon fields \citep[e.g.,][]{ackermann_pwn_search_2011,abdo_velax_2010}. In SNR environments, accelerated particles can also produce gamma-ray emission through leptonic processes or through hadronic interactions with ambient gas, including molecular clouds (MCs) \citep[e.g.,][]{drury_gamma_visibility_1994}. Systems in which an energetic pulsar, its nebula, and a surrounding SNR environment are observed together therefore provide useful laboratories for studying particle acceleration and transport across different spatial scales.

PSR~J1135--6055 was discovered in a \fermi{} blind search and is classified as an energetic gamma-ray-only pulsar \citep{Abdo_2013,2023ApJ...958..191S}. It has $\dot{E}\sim2\times10^{36}$~erg~s$^{-1}$, a characteristic age of $\sim2.3\times10^4$~yr, and a distance of about 2.9~kpc \citep{2017JPlPh..83e6301K}. Chandra observations of the G293.8+0.6 field revealed the pulsar and its nebula in X-rays, while no X-ray emission from the SNR itself was reported \citep{bordas_jetlike_2020}. Deeper Chandra observations show a compact nebula around the pulsar, a trailing tail, and bipolar jet-like structures whose origin may involve either pulsar outflows or the escape of high-energy particles into the ambient magnetic field \citep{bordas_jetlike_2020}.

The radio remnant G293.8+0.6 has also been revisited with the MeerKAT/SMGPS survey. The new radio image resolves a faint, nearly circular shell and a bright central plerionic structure, and the spatially resolved radio spectral index separates the flatter central emission from the steeper shell emission \citep{goedhart_smgps_2024,loru_meerkat_snr_2024}. These X-ray and radio results make the PSR~J1135--6055/G293.8+0.6 field a well-motivated target for a GeV study.

 The paper is organized as follows. Section~\ref{sec:data} describes the \fermi{} data analysis. Section~\ref{sec:interpretation} presents the multiwavelength context and PWN modeling. Section~\ref{sec:discussion} discusses the interpretation and its limitations, and Section~\ref{sec:summary} summarizes the main results.

\section{\texorpdfstring{\fermi}{Fermi-LAT} Analysis and Results}
\label{sec:data}
\subsection{Pulse Profiles and Off-pulse Analysis}
\label{subsec:offpulse_analysis}
 We  first  used the public  15$^\circ$  3PC phase-assigned events for PSR~J1135--6055 from the 3PC data page\footnote{\url{https://fermi.gsfc.nasa.gov/ssc/data/access/lat/3rd_PSR_catalog/3PC_HTML/J1135-6055.html}} \citep[3PC;][]{2023ApJ...958..191S}, covering MJD~54681--58046  (9.21~yr). The file includes the \texttt{PULSE\_PHASE} and \texttt{MODEL\_WEIGHT} columns; the latter gives the pulsar probability weight assigned to each photon.

From the 100~MeV--500~GeV photons within $3^\circ$ of the pulsar, we constructed a 50-bin light curve using the photon-weighting method of \citet{Kerr_2011} and \citet{2019A&A...622A.108B}. The weights depend on the source spectrum, background model, and LAT point-spread function. For a bin containing photons $j$, the statistical uncertainty was estimated as $\sigma = \left(\sum_j w_j^2\right)^{1/2}$ \citep{Abdo_2013}. The expected weighted background level was calculated from the full selected photon sample as $b=\sum_{j} w_{j}(1-w_{j})/n_{\mathrm{bins}}$.

We defined the off-pulse interval from this light curve with a quasi-baseline (QB) procedure \citep{meyer_characterizing_2019}. The QB baseline was estimated from the optimally symmetric low-count part of the weighted phase-histogram distribution, yielding a background-dominated baseline level, $B_0$, and its dispersion, $S_0$. We adopt $\phi=0.14$--0.70 as the off-pulse interval, as marked in Figure~\ref{fig:extended_ephemeris_validation}.
This interval covers 56\% of the rotational phase.  We used the same interval in both source tests below.

 For the first source test,  we selected Pass~8 SOURCE-class events from  1~GeV  to 1~TeV in the phase range $\phi=0.14$--0.70.  The 1~GeV lower bound reduces sensitivity to the broad low-energy LAT point-spread function and low-energy background systematics. We used \texttt{evclass=128}, \texttt{evtype=3}, and the \texttt{P8R3\_SOURCE\_V3} instrument response functions \citep{Atwood_2009}. Events with zenith angles larger than $90^\circ$ were rejected to reduce contamination from the Earth's limb, and good time intervals were selected using \texttt{DATA\_QUAL>0 \&\& LAT\_CONFIG==1}.  This analysis used a $10^\circ \times 10^\circ$ region of interest centered on 4FGL~J1135.2--6055, with  $0.05^\circ$ spatial pixels. The source model was initialized from the 4FGL-DR4 catalog file \texttt{gll\_psc\_v33.fit} \citep{2022ApJS..260...53A,2023arXiv230712546B} and included the standard Galactic interstellar emission model and isotropic diffuse template, \texttt{gll\_iem\_v07.fits} and \texttt{iso\_P8R3\_SOURCE\_V3\_v1.txt}. Likelihood fitting  was performed with \texttt{fermipy} \citep{2017ICRC...35..824W}. Energy dispersion was enabled for source components  and the Galactic interstellar emission, and disabled only for the isotropic component.  The exposure was scaled by the off-pulse phase fraction, 0.56, following the standard treatment for phase-gated LAT analyses.  Following the weighted-likelihood method described in Appendix~B of 4FGL \citep{abdollahi_fermi_2020}, we used \texttt{gtmodel}, \texttt{gteffbkg} (energy-integration factor 2), and \texttt{gtwtsmap} in sequence to generate a position- and energy-dependent weight map for each data span, with $\epsilon=0.03$ for the fractional background-model systematic uncertainty. During the background fit, source normalizations within $5^\circ$, spectral-shape parameters of catalog sources with $\mathrm{TS}\geq25$ within $3^\circ$, the Galactic diffuse normalization and index, and the isotropic normalization were free. In the fits with and without the additional source, we used the same weight map and allowed the same set of background parameters to vary.

 For data spanning MJD~54681--58046 (the 3PC ephemeris validity interval), the weighted off-pulse test of a power-law point source gave $\mathrm{TS}=8.13$, corresponding to $1.71\sigma$ for four degrees of freedom.\footnote{ Here and below, source significances are derived from the $\chi^2$ distribution with the degrees of freedom appropriate to each test and quoted in Gaussian $\sigma$ units, following the LAT catalog convention \citep{abdollahi_fermi_2020}.}

 To increase the photon statistics, we then constructed a phase-connected timing solution for the MJD~54681--60538 interval (16.04~yr). Its construction and validation are described in Appendix~\ref{app:extended_ephemeris}. The solution was tied to the 3PC phase convention over the common interval, allowing us to repeat the same off-pulse weighted-likelihood analysis. The longer data set gives $\mathrm{TS}=15.80$, corresponding to $2.94\sigma$ for four degrees of freedom. This result also remains below the conventional LAT source-detection threshold of $\mathrm{TS}=25$ \citep{abdollahi_fermi_2020}. Thus, even over the currently available 16.04~yr span covered by the extended ephemeris, we find no significant off-pulse gamma-ray excess.

\subsection{Full-phase Analysis}
\label{subsec:high_energy_results}
We performed an additional full-phase analysis, using all pulse phases to improve photon statistics while benefiting from the improved LAT angular resolution at high energies. Similar high-energy LAT selections have been used to study hard GeV emission in PWN and TeV-source fields \citep[e.g.,][]{acero_constraints_2013,zhou_hessj1303_2023,zhou_cta1_2024,gong_j0249_2025,gong_hessj1857_2026}.

\subsubsection{Data Selection and Spatial Models}
For this full-phase analysis, we retained all pulse phases and selected LAT events between 20~GeV and 1~TeV, using the same SOURCE-class event selection, IRFs, diffuse background models, ROI definition, and spatial pixelization as in the off-pulse analysis, except for the phase selection and energy range.  The selected data span MJD~54682.66--61237.25 (17.95~yr). Energy dispersion was enabled for the source components and Galactic diffuse emission and disabled for the isotropic component.  For the morphology comparison, we removed the catalog pulsar and replaced it with a power-law source. We fitted its normalization and photon index, determined the best-fit position, and then tested point, RadialDisk, and RadialGaussian hypotheses.  Starting from the same prefit background model, we reoptimized the normalizations of catalog sources within $3^\circ$ and the diffuse backgrounds for each spatial model, and calculated its 20~GeV--1~TeV TS relative to the reoptimized null model. The source position and photon index obtained from the 20~GeV--1~TeV fit were fixed in the four SED bins.

\subsubsection{Morphological Analysis}
\begin{figure}
    \centering
    \includegraphics[width=\columnwidth]{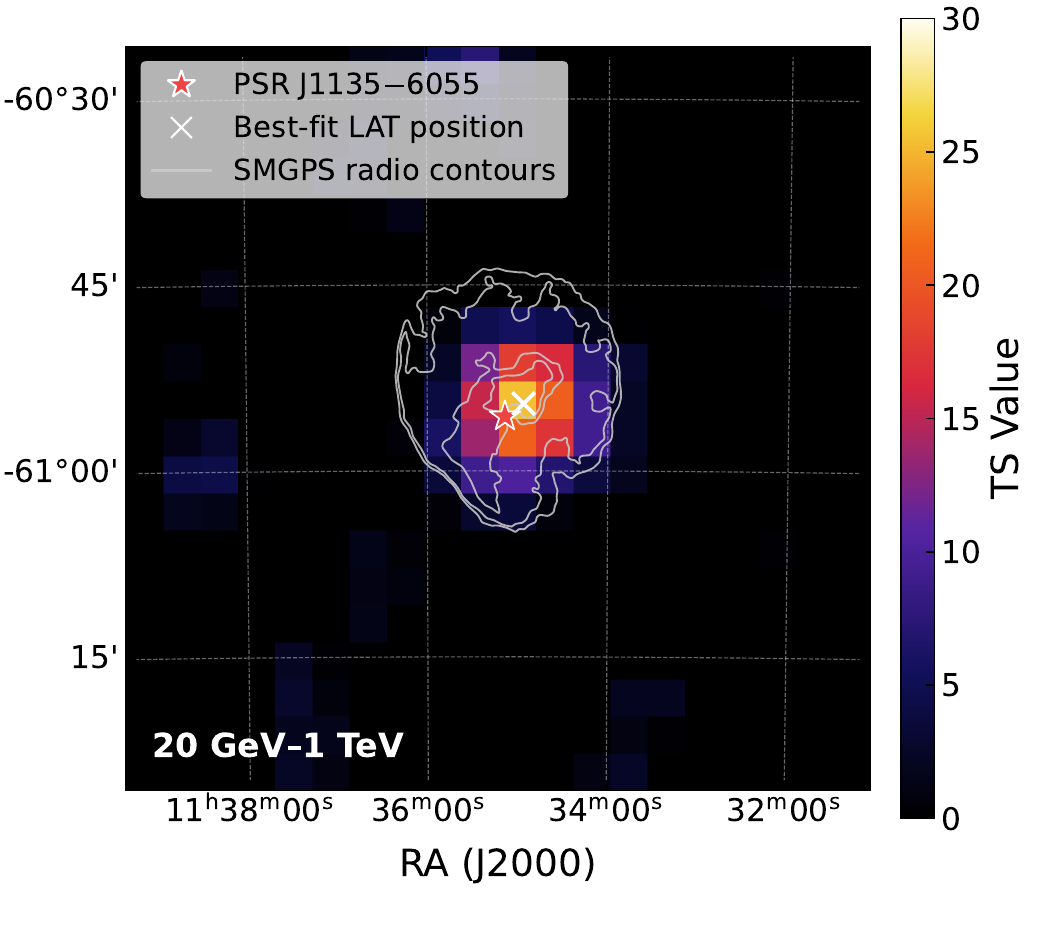}
    \caption{Full-phase 20~GeV--1~TeV TS map before adding the adopted point-source template. The $1^\circ\times1^\circ$ map uses the native $0.05^\circ$ pixels without smoothing. At each pixel, only the normalization of a point test source with $\Gamma=2.0$ was fitted. The red star marks PSR~J1135--6055, the white cross marks the best-fit LAT position, and gray curves show the SMGPS 1.28-GHz radio contours of G293.8+0.6 \citep{goedhart_smgps_2024,goedhart_smgps_dataset_2024,loru_meerkat_snr_2024}.}
    \label{fig:excess_tsmap_20gev}
\end{figure}

Figure~\ref{fig:excess_tsmap_20gev} shows the 20~GeV--1~TeV TS map centered on PSR~J1135--6055.

The point-source fit gives $(\alpha,\delta)=(173.733^\circ,-60.910^\circ)$, $\Gamma=1.63\pm0.33$, and $\mathrm{TS}=26.30$ for four degrees of freedom. The pulsar lies $1.83'$ from the best-fit LAT position, within the 95\% statistical localization radius, $r_{95}\simeq3'$. In Figure~\ref{fig:excess_tsmap_20gev}, the white cross marks the best-fit LAT position, and the red star marks the pulsar.

The disk and Gaussian fits yield $\mathrm{TS}_{\rm ext}=0.24$ and 0.46, with one-sided 95\% limits $R_{\rm disk}<7.4'$ and $\sigma_{\rm G}<3.9'$, respectively; neither indicates significant extension, so we adopt the point-source morphology for the SED extraction. Both upper limits permit a gamma-ray source as broad as the roughly $3'$ radio PWN; the LAT morphology is therefore not demonstrably more compact than the radio nebula.

The gray contours are from the MeerKAT/SMGPS radio-continuum image of G293.8+0.6 \citep{goedhart_smgps_2024,goedhart_smgps_dataset_2024,loru_meerkat_snr_2024}.  The GeV excess lies within the radio extent of G293.8+0.6, close to the central pulsar/PWN region.

\subsubsection{Spectral Analysis}

We extracted the point-source SED in three logarithmically spaced bins between 20~GeV and 500~GeV, with an additional bin from 500~GeV to 1~TeV, allowing the source normalization, the normalizations of catalog sources within $3^\circ$, and the Galactic and isotropic diffuse normalizations to vary in each bin. For a bin with $\mathrm{TS}<4$, we report a one-sided 95\% upper limit. The resulting measurements are listed in Table~\ref{tab:source_components_sed}.

\begin{figure}[!htbp]
    \centering
    \includegraphics[width=\columnwidth]{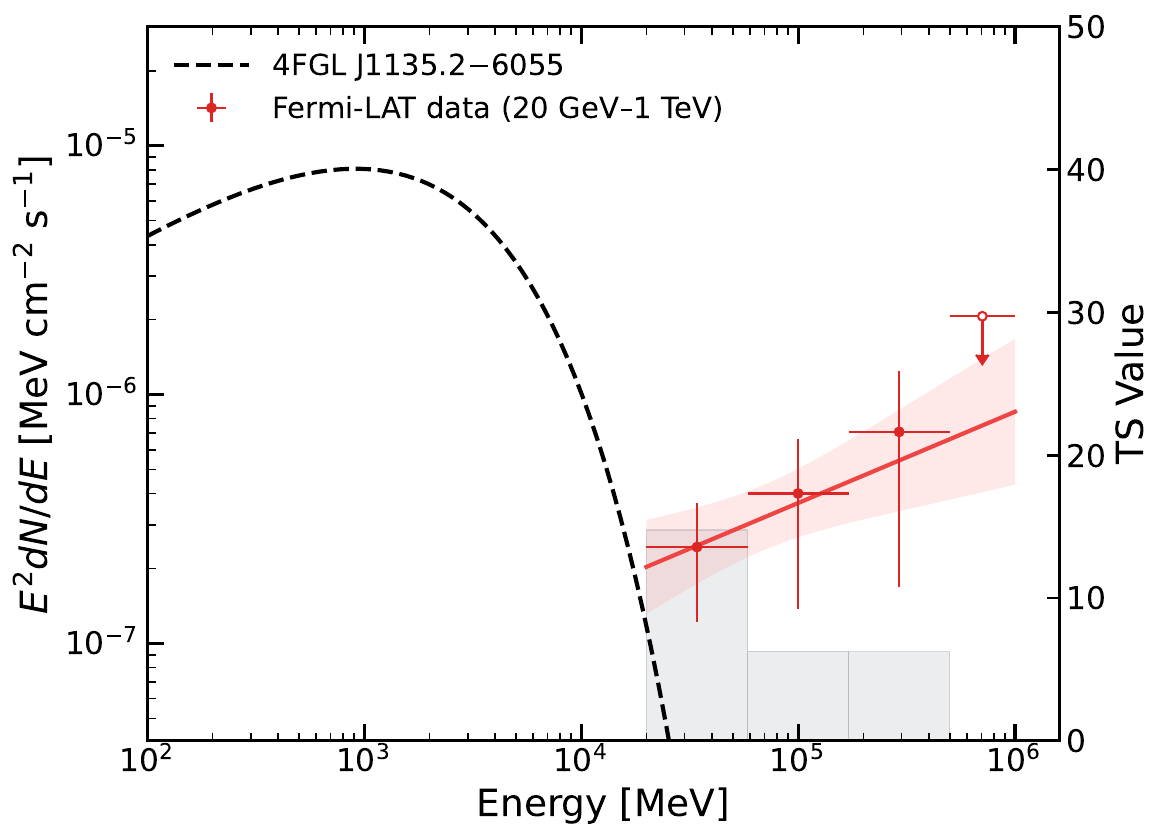}
    \caption{Point-source 20~GeV--1~TeV SED. Red error bars show statistical uncertainties, and a downward arrow denotes a 95\% upper limit. The solid curve and shaded band give the best-fit power law and its statistical uncertainty; gray bars give the TS in each energy bin. The dashed curve is the phase-averaged 4FGL~J1135.2--6055 catalog spectrum \citep{2022ApJS..260...53A,2023arXiv230712546B}.}
    \label{fig:source_components_sed}
\end{figure}

For a point source at these energies, diffuse-model uncertainties are small because of the narrow LAT point-spread function, while the 5--10\% effective-area uncertainty is below the statistical flux errors; Figure~\ref{fig:source_components_sed} and Table~\ref{tab:source_components_sed} therefore show statistical uncertainties only.

\begin{table}[!htbp]
    \centering
    \scriptsize
    \setlength{\tabcolsep}{0.8pt}
    \renewcommand{\arraystretch}{1.12}
    \caption{Point-source 20~GeV--1~TeV SED measurements. Flux uncertainties are statistical only, and bins with $\mathrm{TS}<4$ are reported as 95\% upper limits.}
    \label{tab:source_components_sed}
    
    \begin{tabular}{ccc}
        \toprule
        Energy range & TS & $E^2dN/dE$ \\
        (GeV) & & ($10^{-7}$~MeV~cm$^{-2}$~s$^{-1}$) \\
        \midrule
        20.00--58.48 & 14.8 & $2.4 \pm 1.2$ \\
        58.48--171.00 & 6.3 & $4.0 \pm 2.6$ \\
        171.00--500.00 & 6.2 & $7.1 \pm 5.4$ \\
        500.00--1000.00 & 0.0 & $<21$ \\
        \bottomrule
    \end{tabular}
\end{table}

Because this analysis uses all pulse phases and the magnetospheric and putative nebular components are spatially unresolved by the LAT, the phase-averaged 4FGL pulsar spectrum in Figure~\ref{fig:source_components_sed} may contain magnetospheric emission in the lowest-energy SED bin.

\section{Multiwavelength Context and PWN Modeling}
\label{sec:interpretation}

\subsection{Multiwavelength Context}
\label{subsec:pwn_candidate}
Based on the MeerKAT/SMGPS radio observations, \citet{loru_meerkat_snr_2024} describe G293.8+0.6 as a composite SNR with a bright central plerionic component and a faint, nearly circular shell. Aperture photometry of the 1.284-GHz SMGPS image \citep{goedhart_smgps_2024,goedhart_smgps_dataset_2024}, using a $3'$ source aperture and a $3.5'$--$5.5'$ background annulus, gives $S_\nu=0.350\pm0.028$~Jy for the central PWN. The quoted uncertainty combines the photometric error with the 8\% calibration uncertainty adopted for SMGPS flux-density measurements by \citet{loru_meerkat_snr_2024}.

The radio part of the PWN fit includes the representative central-PWN spectral index $\alpha_R\simeq-0.21$ reported by \citet{loru_meerkat_snr_2024}, defined by $F_\nu\propto\nu^{\alpha_R}$. The model index is evaluated over the corresponding 0.155--1.284-GHz interval, while the 1.284-GHz flux density provides the normalization. The normalization and representative index form an approximate constraint because their angular resolutions and apertures are not strictly matched.

 At X-ray energies, Chandra observations reveal the pulsar, a compact nebula, and jet-like structures \citep{bordas_jetlike_2020}. We use the summed 1-keV emission of Jet-1 and Jet-2 as the X-ray constraint. Their reported normalizations give $E^2dN/dE=(3.70\pm0.21)\times10^{-14}$~erg~cm$^{-2}$~s$^{-1}$. We approximate their summed 0.5--7~keV spectrum by a single power law with photon index $\Gamma_X=1.77\pm0.07$. The summed normalization and slope constrain the independent jet model.

At TeV energies, the field is covered by the H.E.S.S. Galactic Plane Survey, but no HGPS catalog source is coincident with PSR~J1135--6055 or G293.8+0.6 \citep{hess_collaboration_hess_2018}. The source is outside the HAWC and LHAASO survey regions \citep{albert_3hwc_2020,cao_first_2024}. At the pulsar position, the public HGPS $0.1^\circ$ map gives a 95\% confidence-level upper limit of $F(>1~{\rm TeV})\mathrel{<}2.21\times10^{-13}$~ph~cm$^{-2}$~s$^{-1}$ for a photon index of 2.3.

 At $\sim2\times10^4$~yr, the PWN is expected to have passed through the SNR reverse shock and may have been compressed, displaced, or mixed with shocked ejecta \citep{blondin_pwn_evolved_2001,gelfand_dynamical_2009,zhu_multiband_2018}. This evolution and the LAT excess position motivate an evolved-PWN interpretation. We next model the multiwavelength emission with a one-zone leptonic model.

\subsection{One-zone Leptonic Modeling}
\label{subsec:pwn_model}
We use a one-zone leptonic model to calculate the synchrotron and inverse-Compton emission from the PWN. We adopt a source distance $d=2.9$~kpc \citep{saz_parkinson_eight_2010,bordas_jetlike_2020}.

\begin{figure*}[!t]
    \centering
    \includegraphics[width=\textwidth]{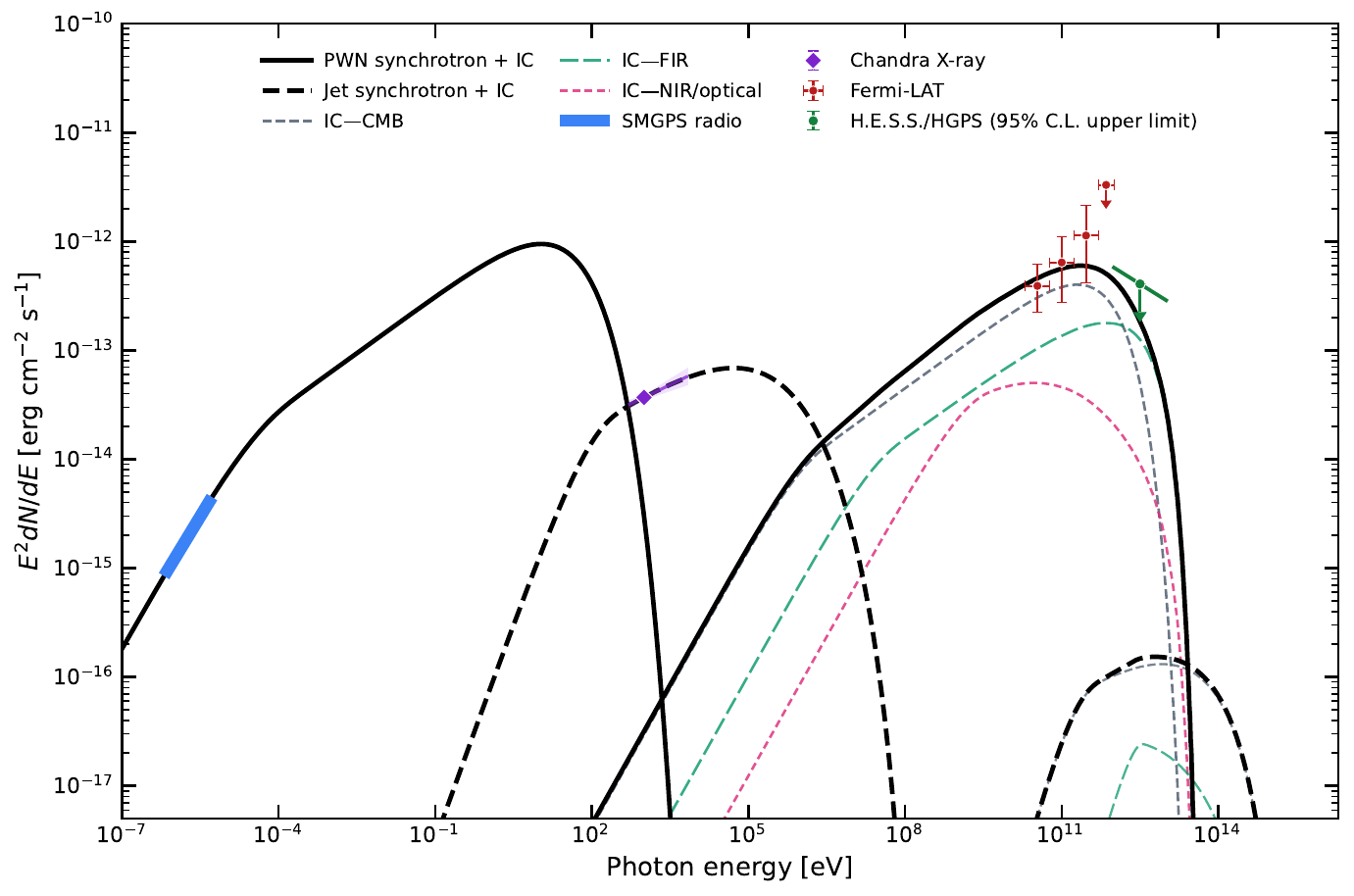}
    \caption{Modeling of the multiwavelength data with a one-zone PWN model and an independent jet component. The black solid and dashed curves show the total PWN and jet emission, respectively, including both synchrotron and inverse-Compton radiation; thinner dashed curves show the seed-field IC contributions. The multiwavelength constraints are from SMGPS \citep{goedhart_smgps_2024,goedhart_smgps_dataset_2024,loru_meerkat_snr_2024}, the summed Chandra spectrum \citep{bordas_jetlike_2020}, the \fermi{} analysis presented here, and HGPS \citep{hess_collaboration_hess_2018}. The blue radio segment uses the representative index $\alpha_R\simeq-0.21$ from \citet{loru_meerkat_snr_2024} over 0.155--1.284~GHz, normalized to our 1.284-GHz flux density. The HGPS upper-limit segment follows $\nu F_\nu\propto E^{-0.3}$.}
    \label{fig:pwn_broadband_sed}
\end{figure*}

The electron--positron energy distribution, $N(\gamma)\equiv dN_e/d\gamma$, is assumed to follow a broken power law with a super-exponential cutoff. For $\gamma_{\min}\leq\gamma\leq\gamma_{\max}$,

\begin{equation}
\begin{aligned}
N(\gamma)
&=N_0\exp\!\left[-\left(\frac{\gamma}{\gamma_{\rm cut}}\right)^{\beta}\right]\\
&\quad\times
\begin{cases}
\left(\gamma/\gamma_{\rm b}\right)^{-\alpha_1}
 & \gamma\leq\gamma_{\rm b}\\
\left(\gamma/\gamma_{\rm b}\right)^{-\alpha_2}
 & \gamma>\gamma_{\rm b}
\end{cases}.
\end{aligned}
\label{eq:pwn_electron_spectrum}
\end{equation}

Here $\gamma$ is the lepton Lorentz factor, $\alpha_1$ and $\alpha_2$ are the spectral indices below and above the break $\gamma_{\rm b}$, $\gamma_{\rm cut}$ is the cutoff Lorentz factor, and $\beta$ is the cutoff exponent. The normalization $N_0$ is fixed by the total kinetic energy $W_e$ in the modeled electron--positron population:

\begin{equation}
W_e=\int_{\gamma_{\min}}^{\gamma_{\max}}
(\gamma-1)m_ec^2N(\gamma)\,d\gamma.
\label{eq:pwn_energy_normalization}
\end{equation}

Here $m_e$ is the electron mass and $c$ is the speed of light.

Inverse-Compton emission is calculated with the full Klein--Nishina cross section \citep{blumenthal_gould_1970}.

The radiative calculation includes synchrotron emission and inverse-Compton scattering on the CMB and on far-infrared and near-infrared/optical photon fields based on a GALPROP model of the interstellar radiation field (ISRF) \citep{porter_3d_isrf_2017}.

The seed fields are represented by blackbody components with $(T,u)=(2.73~{\rm K},0.25~{\rm eV\,cm^{-3}})$ for the CMB, \textcolor{black}{$(66.84~{\rm K},0.27~{\rm eV\,cm^{-3}})$ for the FIR field}, and $(5000~{\rm K},1.26~{\rm eV\,cm^{-3}})$ for the near-infrared/optical field. \textcolor{black}{We adopt the R12 ISRF model described by \citet{porter_3d_isrf_2017} and evaluate its spectral energy density at the three-dimensional Galactic position of PSR~J1135--6055, adopting $d=2.9$~kpc. The FIR and NIR/optical parameters were obtained by representing the corresponding portions of this source-position spectrum with diluted blackbody components. These parameters define the soft-photon fields used in the inverse-Compton calculation.}

\citet{bordas_jetlike_2020} tentatively associate the compact emission ahead of the pulsar with a bow-shock region and discuss two possible explanations for the lateral structures: true pulsar jets or synchrotron features produced by high-energy particles escaping from the PWN into the ambient magnetic field. Because these structures are morphologically distinct and their origin remains uncertain, we model their summed emission independently of the PWN calculation. To represent this separate component, we adopt a power-law electron distribution with an exponential cutoff and use \texttt{naima} \citep{zabalza_naima_2015} to calculate the resulting synchrotron and inverse-Compton emission. Following the reference setup in that study, we set $B=50~\mu$G and assume $E_{\rm cut}=256$~TeV, with the latter consistent with the X-ray-emission and particle-confinement constraints derived there. The adopted parameters and fit results are listed in Table~\ref{tab:pwn_model_params}. For consistency, the PWN and jet inverse-Compton calculations use the same soft-photon fields.

In this picture, the GeV photons are produced mainly by inverse-Compton scattering from relativistic electrons and positrons, while the same lepton population produces synchrotron emission constrained by the radio data.

The gamma-ray terms use the LAT per-bin likelihood profiles and the HGPS aperture-integrated upper limit above 1~TeV.

Table~\ref{tab:pwn_model_params} lists the principal model parameters. We fitted $\alpha_1$, $\alpha_2$, $B$, and $W_e$, while $d$, the target photon fields, $\gamma_{\min}$, $\gamma_{\max}$, $\gamma_{\rm b}$, $\gamma_{\rm cut}$, and $\beta$ were fixed. The break and cutoff parameters are fixed at the values listed in Table~\ref{tab:pwn_model_params}. The parameters of the independent jet component are listed in the same table.

\begin{table}[!htbp]
    \centering
    \scriptsize
    \setlength{\tabcolsep}{1.5pt}
    \setlength{\lightrulewidth}{0.10em}
    \setlength{\heavyrulewidth}{0.14em}
    \renewcommand{\arraystretch}{0.90}
    \caption{Parameters of the one-zone PWN model and independent jet component. The Origin column indicates whether each value was adopted from published measurements or models, assumed, fitted, or derived. For the PWN, $W_e$ is integrated over $\gamma_{\min}\leq\gamma\leq\gamma_{\max}$.}
    \label{tab:pwn_model_params}
    
    \begin{tabular}{@{}>{\raggedright\arraybackslash}p{0.34\columnwidth}
                        >{\centering\arraybackslash}p{0.41\columnwidth}
                        >{\centering\arraybackslash}p{0.18\columnwidth}@{}}
        \toprule
        Parameter & Value & Origin \\
        \midrule[\heavyrulewidth]
        \multicolumn{3}{l}{PWN component} \\
        \midrule[\heavyrulewidth]
        \multicolumn{3}{l}{Source} \\
        \midrule
        $d$ (kpc) & 2.9 & Adopted \\
        \midrule
        \multicolumn{3}{l}{Environment} \\
        \midrule
        $B$ ($\mu$G) & \leavevmode4.81 & Fitted \\
        CMB $(T,u)$ & (2.73~K, 0.25~eV~cm$^{-3}$) & Adopted \\
        FIR $(T,u)$ & (66.84~K, 0.27~eV~cm$^{-3}$) & Adopted \\
        NIR/optical $(T,u)$ & (5000~K, 1.26~eV~cm$^{-3}$) & Adopted \\
        \midrule
        \multicolumn{3}{l}{Electron spectrum} \\
        \midrule
        $\alpha_1$ & 1.29 & Fitted \\
        $\alpha_2$ & 2.30 & Fitted \\
        $\gamma_{\min}$ & 1 & Assumed \\
        $\gamma_{\max}$ & $1.00\times10^{10}$ & Assumed \\
        $\gamma_{\rm b}$ & $3.00\times10^{4}$ (15.3~GeV) & Assumed \\
        $\gamma_{\rm cut}$ & \leavevmode$2.50\times10^{7}$ (12.8~TeV) & Assumed \\
        $\beta$ & 2 & Assumed \\
        \leavevmode$W_e$ (erg) & $2.03\times10^{47}$ & \leavevmode{}Fitted \\
        \midrule[\heavyrulewidth]
        \multicolumn{3}{l}{Jet component} \\
        \midrule[\heavyrulewidth]
        $\alpha$ & 2.38 & Fitted \\
        $E_{\rm cut}$ (TeV) & 256 & Assumed \\
        $B$ ($\mu$G) & 50 & Assumed \\
        $W_e(\gamma>10^7)$ (erg) & $1.88\times10^{42}$ & Derived \\
        \bottomrule
    \end{tabular}
\end{table}

With the reasonable parameter values listed in Table~\ref{tab:pwn_model_params}, \textcolor{black}{the PWN synchrotron spectrum reproduces the 1.284-GHz radio normalization, and a power-law fit over the adopted radio interval gives $\alpha_R=-0.21$.} The independent jet synchrotron spectrum reproduces both the summed 1-keV X-ray normalization and the adopted 0.5--7~keV slope. The PWN inverse-Compton component reproduces the LAT measurements and remains consistent with the HGPS upper limit, as shown in Figure~\ref{fig:pwn_broadband_sed}.

\section{Discussion}
\label{sec:discussion}
 The radio and X-ray observations summarized in Section~\ref{subsec:pwn_candidate} establish the presence of a PWN within the SNR G293.8+0.6. Chandra observations reveal a compact nebula around the pulsar, a trailing tail, and bipolar jet-like structures \citep{bordas_jetlike_2020}. The X-ray PWN overlaps the eastern side of the radio plerion \citep{loru_meerkat_snr_2024}. The spin-down luminosity of PSR~J1135--6055, $\dot{E}\sim2\times10^{36}$~erg~s$^{-1}$, is sufficient to power the nebula \citep{2017JPlPh..83e6301K}.

The LAT point source is positionally consistent with PSR~J1135--6055 and the PWN observed at radio and X-ray wavelengths.  Its photon index is $\Gamma=1.63\pm0.33$. Although it is unresolved, the LAT point source can be as extended as the radio PWN. Hard gamma-ray spectra have also been reported for other LAT PWNe and PWN candidates \citep{acero_constraints_2013,the_fermi-lat_collaboration_systematic_2025}. Because the high-energy analysis uses all pulse phases and the magnetospheric and putative nebular components are spatially unresolved by the LAT, a magnetospheric contribution cannot be excluded; the phase-averaged 4FGL pulsar spectrum in Figure~\ref{fig:source_components_sed} may contribute particularly to the lowest-energy SED bin. Although the weighted off-pulse analysis does not independently detect this source,  these positional and spectral properties,  together with the broadband fit in Section~\ref{subsec:pwn_model},  support its interpretation as a GeV PWN candidate.

 We considered halo and hadronic interpretations. For a halo interpretation, the absence of significant LAT extension provides no evidence for a resolved GeV halo.  A hadronic interpretation would require relativistic protons interacting with sufficiently dense ambient gas. To examine this possibility, we inspected the SGPS H~I and Mopra CO data between  $-24$ and $-18$~km~s$^{-1}$, guided by the H~I emission reported along the southeastern radio edge of G293.8+0.6 \citep{loru_meerkat_snr_2024,mcclure_griffiths_sgps_2005,cubuk_mopra_dr4_2023}.  These data do not reveal a reliable compact gas counterpart at the LAT position. A likelihood test using a fixed H~I template constructed from the same velocity interval, either in place of the point source or as an additional spatial component, did not improve the fit. Thus, the available gas observations do not provide clear support for a hadronic origin, although unresolved or poorly traced target gas cannot be excluded.

\section{Summary}
\label{sec:summary}
We have analyzed the GeV emission around PSR~J1135--6055 using the phase-selected and full-phase LAT analyses described above.  For the MJD~54681--60538 interval, the weighted 1~GeV--1~TeV off-pulse point-source test gives $\mathrm{TS}=15.80$, below the LAT source-detection threshold.

 By contrast, the 20~GeV--1~TeV LAT analysis yields a point source with $\mathrm{TS}=26.30$ and $\Gamma=1.63\pm0.33$. The best-fit LAT position is consistent with PSR~J1135--6055 within the 95\% statistical localization region, and neither the disk nor Gaussian morphology shows significant extension. The extension limits allow a gamma-ray source as broad as the roughly $3'$ radio PWN.

 A one-zone leptonic model together with an independent jet component constrained by the summed X-ray spectrum, reproduces the multiwavelength observations and remains consistent with the HGPS limit. The available gas data provide no clear support for a hadronic origin.  Together, the positional agreement with PSR~J1135--6055 and its radio/X-ray PWN, the hard LAT spectrum, and the broadband model support interpreting the source as a GeV PWN candidate, although its physical origin remains unconfirmed.

 Future CTAO-South observations could search for a TeV counterpart to the LAT point source. A detection or a more stringent upper limit would test the predicted inverse-Compton emission and constrain the maximum energy of the electron population \citep{cta_science_2019}.

\appendix
\section{Extended Phase-connected Timing Solution}
\label{app:extended_ephemeris}
\setcounter{figure}{0}
\renewcommand{\thefigure}{\thesection.\arabic{figure}}
\providecommand{\theHfigure}{}
\renewcommand{\theHfigure}{appendix.\thesection.\arabic{figure}}
 Following standard LAT pulsar-timing procedures \citep{ray_precise_2011,kerr_timing_2015}, we used PINT \citep{2021ApJ...911...45L} to extend the published 3PC phase-connected timing solution \citep{2023ApJ...958..191S}. Using LAT gamma-ray times of arrival (TOAs), we fitted $F_0$--$F_2$ together with five harmonically related WAVEX sine--cosine pairs to describe the low-frequency timing noise, while retaining the published glitch parameters. The resulting ephemeris is phase connected over the MJD~54681--60538 interval.

\begin{figure}[!ht]
    \centering
    \includegraphics[width=0.82\textwidth,height=0.50\textheight,keepaspectratio]{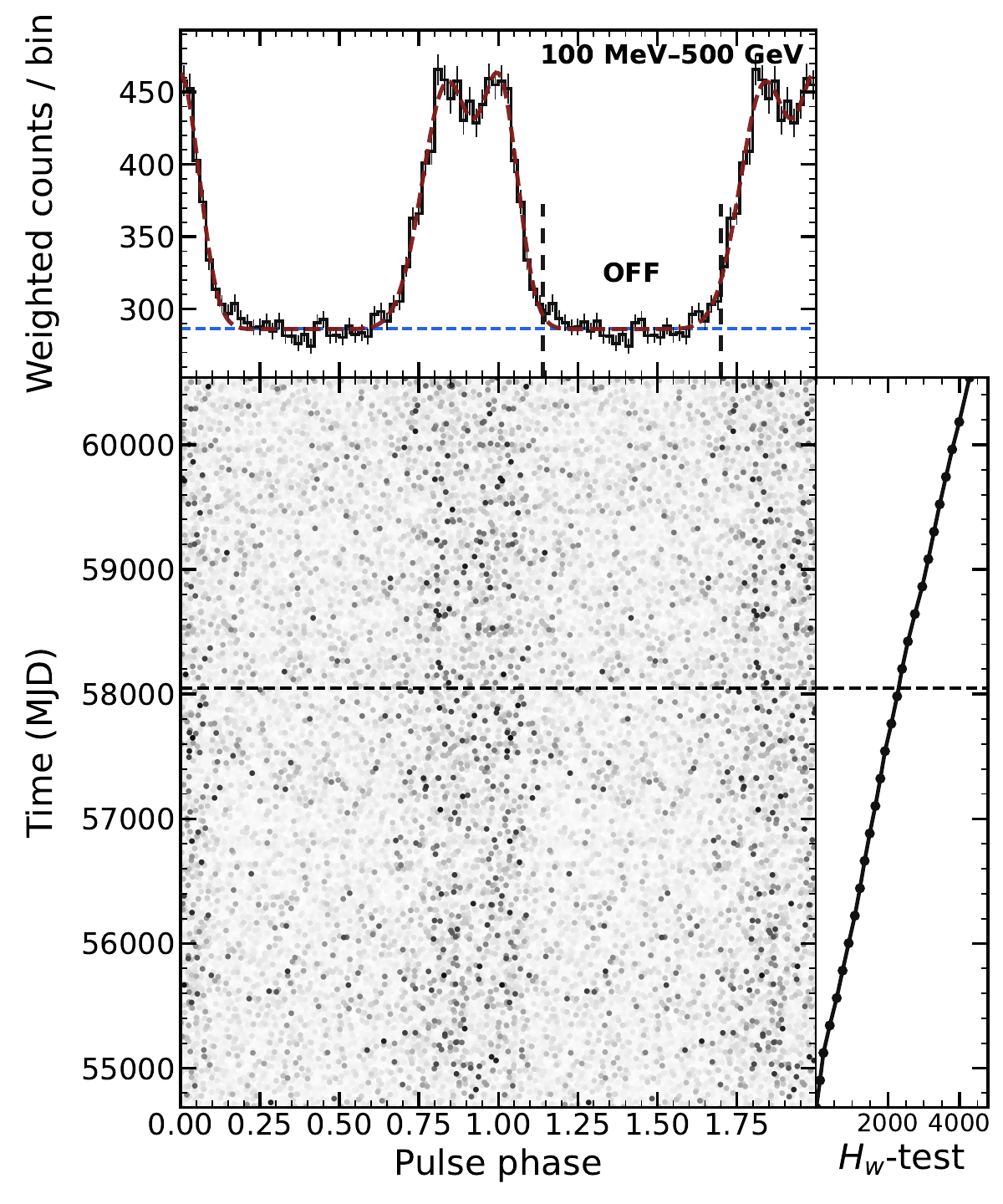}
    \caption{ Weighted pulse profile, phase--time diagram, and cumulative weighted $H$-test curve for PSR~J1135--6055  obtained with the extended phase-connected timing solution for the MJD~54681--60538 interval.  The upper panel shows the  100~MeV--500~GeV  weighted pulse profile over two rotations. The lower-left panel shows pulse phase versus MJD, with grayscale indicating photon weight, and the lower-right panel shows the cumulative weighted $H$ statistic in 220-day steps. The red dashed curve shows the result of an empirical double-Gaussian fit to the weighted pulse profile, the blue dashed line marks the expected weighted background, and the black dashed lines mark the off-pulse interval $\phi=0.14$--0.70 for reference. The black horizontal dashed line in each lower panel marks MJD~58046, the end of the 3PC timing interval. The phase--time diagram shows that the pulse phase remains stable throughout the validity interval of the extended timing solution.}
    \label{fig:extended_ephemeris_validation}
\end{figure}

 The solution could not be reliably extrapolated beyond MJD~60538, possibly because of an unmodeled glitch or other timing irregularity. We therefore restrict its validity to the MJD~54681--60538 interval.  The ephemeris file \texttt{J1135-6055\_ephemeris.par} is provided as supporting data with this article.

\FloatBarrier
\begin{acknowledgments}
This work made use of publicly available \fermi{} data obtained through the Fermi Science Support Center (FSSC) at NASA Goddard Space Flight Center.  We also used MeerKAT/SMGPS DR1 radio-continuum data \citep{goedhart_smgps_dataset_2024}, SGPS H~I data \citep{mcclure_griffiths_sgps_2005}, Mopra Southern Galactic Plane CO Survey DR4 data \citep{cubuk_mopra_dr4_2023}, and public HGPS map products \citep{hess_collaboration_hess_2018}. The MeerKAT telescope is operated by the South African Radio Astronomy Observatory, which is a facility of the National Research Foundation, an agency of the Department of Science and Innovation. We thank Dr. Sara Loru for helpful correspondence regarding the radio observations of G293.8+0.6.

This work was partially supported by the National Natural Science Foundation of China (12233006, 12063004, 12393852, 12103046, 12503016, 12588202), by Yunnan Fundamental Research Projects (grant No. 202501AS070068), and by the Project of Yunnan Provincial Department of Education Science Research Fund (2026Y0173).
\end{acknowledgments}

\bibliography{AAS77860R3}
\bibliographystyle{AAS77860R3}

\end{document}